\documentclass[prb,twocolumn]{revtex4-1} 

\usepackage{comment}
\usepackage{amsmath} 
\usepackage{amssymb}

\usepackage{graphicx} 

\usepackage{hyperref}
\hypersetup{colorlinks=true,linkcolor=red,citecolor=blue,urlcolor=black,bookmarksdepth=subsubsection} 

\begin{document}

\title{Revisiting the round bottom flask rainbow experiment}
\author{Markus Selmke, Sarah Selmke}
\affiliation{*Universit\"at Leipzig, 04103 Leipzig, Germany}
\email{markus.selmke@gmx.de}
\homepage{http://photonicsdesign.jimdo.com}
\date{\today}

\begin{abstract}
A popular demonstration experiment in optics uses a round-bottom flask filled with water to project a circular rainbow on a screen with a hole through which the flask is illuminated. We show how the vessel's wall shifts the second-order and first-order bows towards each other and consequentially narrows down Alexander's dark band. We address the challenge this introduces in producing Alexander's dark band, and explain the importance of a sufficient distance of the flask to the screen. The wall-effect also introduces a splitting of the bows which can easily be misinterpreted.
\end{abstract}


\maketitle 




\section{Introduction}
Rainbows present themselves to the observer as a fascinating and varied phenomenon. In geometrical optics, the explanation involves the discussion of light rays through a spherical drop and their bunching for particular deflection angles corresponding to the first and second-order rainbows, respectively. Several exhaustive, yet well accessible reviews of the theoretical concepts are available.\cite{Nussenzveig1977,Walker1975} A review on recent advances in the field gives further details on matters such as drop-shape influences and reflection or reflected bows.\cite{Hau§mann2016} More elementary introductions focusing on the geometrical optics aspects are also abundantly available.\cite{Whitaker1974,Hendry2003,Casini2012}

\begin{figure}[bth]
\begin{center}\includegraphics [width=1.0\columnwidth]{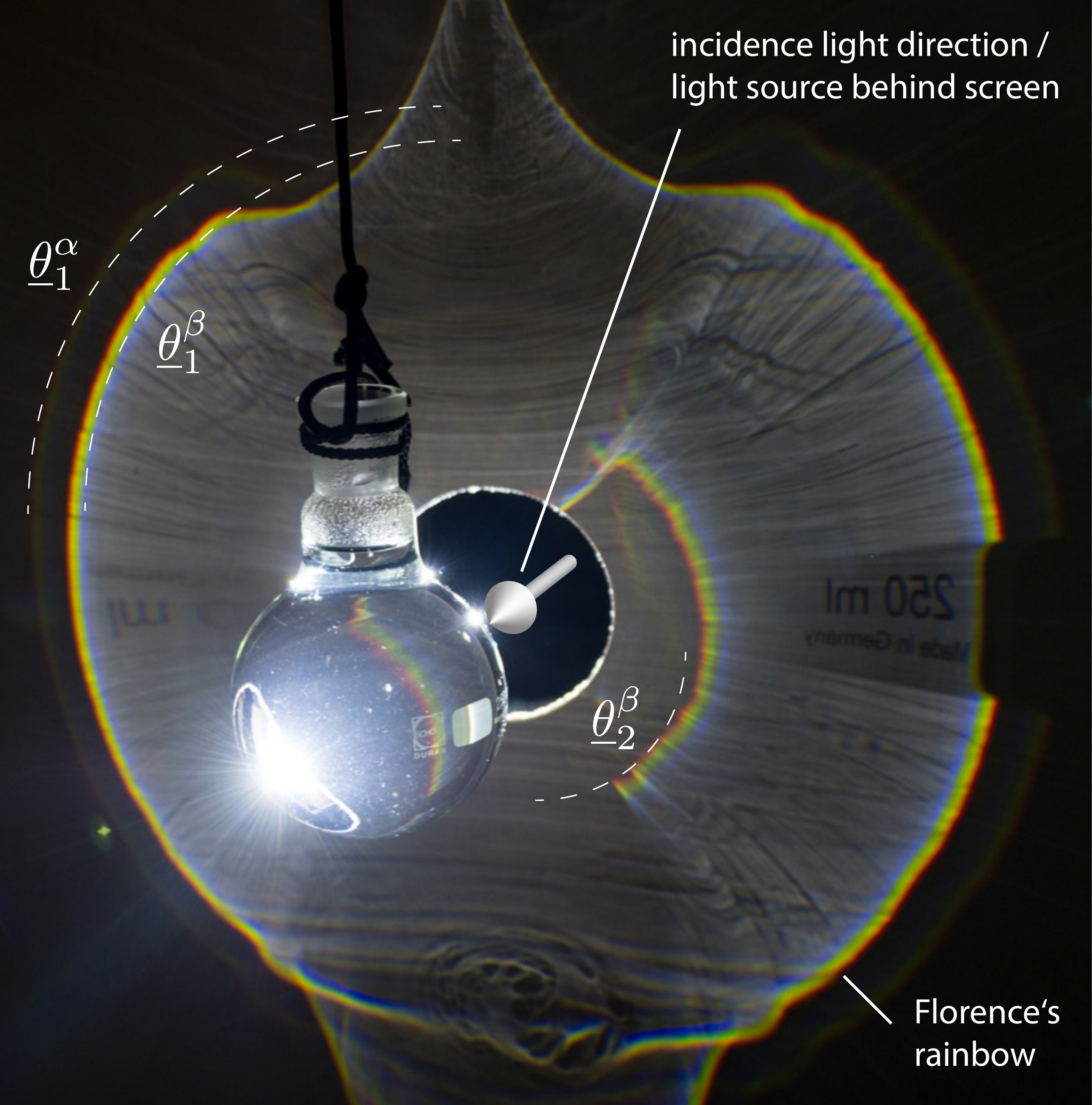}\end{center} %
\caption{Rainbow demonstration experiment: A water-filled round-bottomed flask (image: $250\,\rm mL$ flask, $\varnothing=8.5\,\rm cm$, wall-thickness: $1.7\,\rm mm$) is illuminated by parallel white light through a hole (of slightly larger diameter than the flask) in a white screen. A circular rainbow appears on the screen towards the light source. For the given screen distance of $l\approx 20\,\rm cm$ from the flask, a bright band between the first and second order rainbows is visible instead of Alexander's dark band. Further features are described in this article.\label{Fig_Exp}} 
\end{figure}

The use of a round-bottomed (Florence) flask / globe filled with water as a model raindrop has long been a way of experimentally studying the rainbow formation. The practice of such experiments goes back at least to Theodoric of Freiburg in the 14th century and later Descartes (\textit{L\'es M\'et\'eores}, 1637). As an educational demonstration experiment, also called Florence's rainbow (see Fig.\ \ref{Fig_Exp}), it has been around since at least 1892,\cite{Greenslade2982,Johnson} and was described in detail in the book of Minnaert\cite{Minnaert1993} and other books\cite{Ticker,Bohren1980} as well as in various online resources.\cite{Calvert,Harvard}

Typically, the effects of the finite wall-thickness are ignored for the benefit of a clearer exposition and analogy. However, it is the aim of this note to draw attention to several experimental observations in this classic experiment which are associated with precisely this detail. 

To do so, we will draw on previous work from a domain of metrology, specifically liquid cell refractometry,\cite{Hattori1998,Wang2014} and several studies on the topic of coated sphere rainbow scattering,\cite{Lock1994,Lock2012} and apply it to the experiment. We will provide accessible derivations of all relevant results. The additional refraction taking place at the wall of the vessel changes the observed rainbow angles. Also, due to the more complex geometry, an increased variety of ray paths exists. This introduces an easily misinterpreted splitting of the ordinary rainbows. 


The second pitfall addressed in this note stems from the near-field characteristics of the experiment.\cite{Calvert} For practical reasons, the projection screen is often placed rather close to the illuminated flask,\cite{Greenslade2982,Johnson,Minnaert1993,Ticker,Bohren1980,Calvert,Harvard} cf.\ Fig.\ \ref{Fig_Exp}. The reason being that this ensures that the first-order rainbow, which makes an angle of $\underline{\theta}_1^\beta \sim 42^{\circ}$ to the optical axis, is bright and still fits a convenient screen size. However, only for a sufficiently large critical distance of the projection screen to the flask will Alexander's dark band be visible between the first and second order rainbows. These constrains become even more severe as the vessel wall-thickness increases.

While demonstration experiments which circumvent these issues exist, e.g.\ considering artificial rainbows from acrylic spheres/discs\cite{Harvard,Casini2012,Selmke2015} or from small suspended drops of liquid,\cite{Walker1975,Walker1980,Tammer1998} the Florence flask experiment remains to be an intuitively accessible demonstration experiment. It may therefore appear worthwhile to caution teachers and students alike of some of the subtleties involved in this experiment, either for the purpose of prevention of potential misinterpretation or as a guide for a more detailed treatment in special seminars or homework assignments.

\section{Rainbows of spherical flasks}
\begin{figure}[tbh]
\begin{center}\includegraphics [width=1.0\columnwidth]{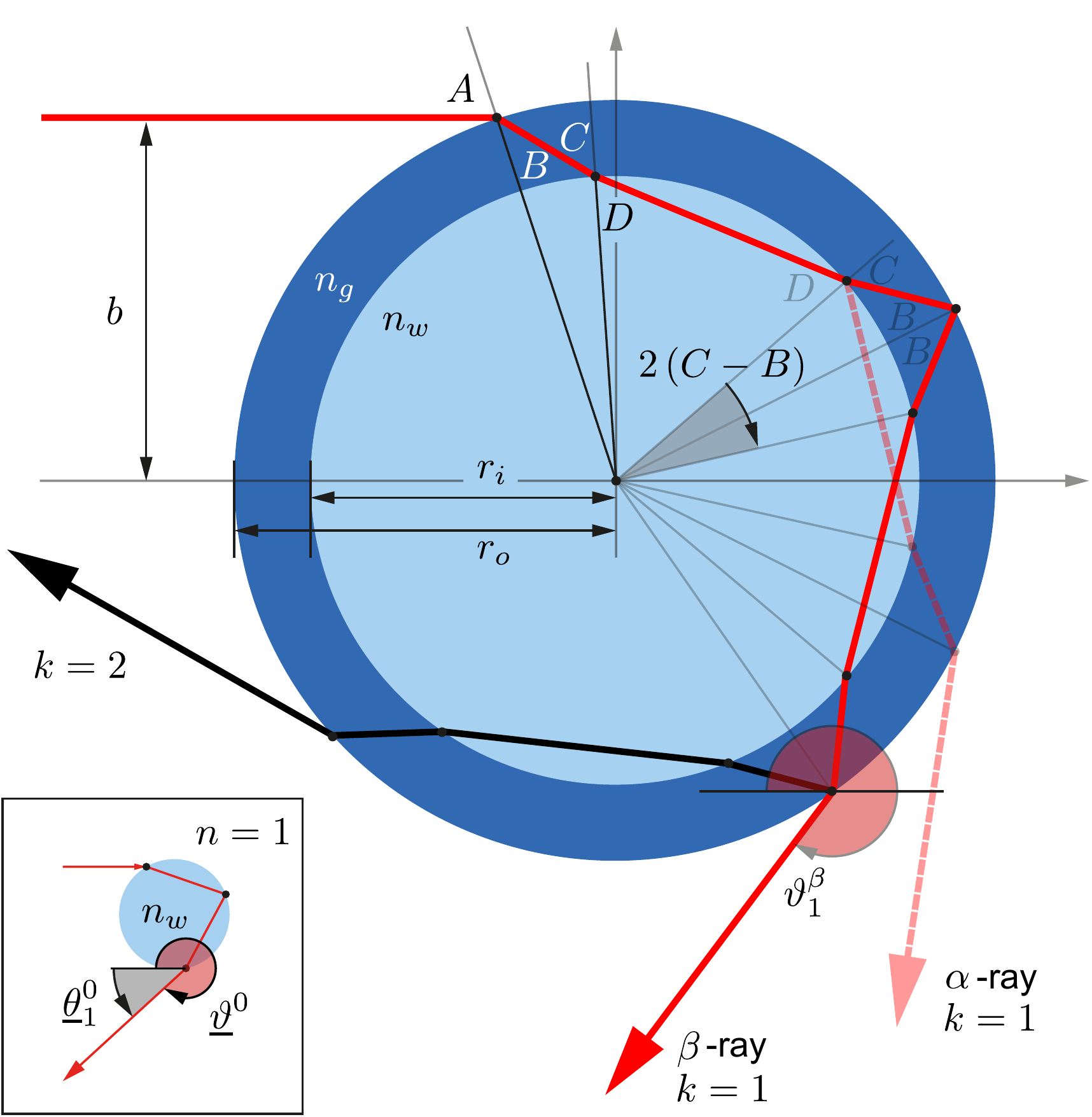}\end{center} 
\caption{Geometry of refraction for a $\varnothing=2r_o$-flask with wall-thickness $\delta=r_o-r_i$. The refractive indices of the wall material, the contained water and air are $n_g$, $n_w$ and $n=1$, respectively. The inset (bottom left) shows the situation of a spherical raindrop (without coating). The drawn $\beta$-ray corresponds to the Cartesian $k=1$ ray (ray of minimum deviation) in the sketch such that $A=\underline{A}$. This is not simultaneously true for the $\alpha$-ray, nor is it the case for $k=2$.\label{Fig_Sketch}}
\end{figure}

The geometry of the relevant rays discussed here is shown in Fig.\ \ref{Fig_Sketch}. Upon entry, a ray's first encounter with the model raindrop, i.e. the water-filled sphere, will be the glass-air interface which is spherical and described by the outer radius $r_o$. A ray with a distance $b$ to an axis parallel to the incident direction and passing through the center of the sphere will make an angle $A=\arcsin\left(b/r_o\right)$ to the interface normal. By Snell's law, the first refraction results in the internal angle $B=\arcsin\left(\sin\left(A\right)/n_g\right)$. The ray continues through the glass medium until it encounters the spherical glass-water interface described by an inner radius $r_i$. The angle $C\left(A,r_i,r_o\right)$ may be obtained by elementary geometric means and a derivation is given in Appendix \ref{AppendixAngleC}. The refracted angle inside the water is found again by applying Snell's law, $D=\arcsin\left(n_g \sin\left(C\right)/n_w\right)$. From this point on, several different ray paths are possible. An extensive collection of possible paths are illustrated for instance in Fig.\ 2 of Ref \cite{Lock2012}. We first discuss the two variants for the 1st-order rainbow. They have been coined the $\alpha$-ray, describing the ray path for an internal reflection at the inner low-contrast ($n_w$ to $n_g$) water-glass interface, and the $\beta$-ray, describing the ray trajectory that results from a reflection at the outer high-contrast ($n_g$ to $n=1$) glass-air interface. Both of these rays are depicted in Fig.\ \ref{Fig_Sketch}. Due to the higher refractive index contrast, the $\beta$-rays will be responsible for the clearly visible first-order rainbow and corresponds to the experimental rainbow observed in the context of this experiment. It is this ray which is then taken as a proxy for the no-wall rays discussed in the theory of the natural rainbow. For the coated sphere we consider here, we will see that both rays are actually observable and that both show \textit{deviations from the rainbow theory reference values} $\underline{\vartheta}^0_k$ (and $\underline{\theta}^0_k$, e.g.\ $\underline{\theta}^0_1\approx 42^{\circ}$).

To show this, one proceeds in the same vein as in the original rainbow theory,\cite{Walker1975,Whitaker1974,Casini2012,Adam2002} and determines the overall deviation angle of the ray by considering the sum of the individual deviations by all involved reflections and refractions. Referring to Fig.\ \ref{Fig_Sketch} one finds $\vartheta^\beta_k\left(A\right)=\left(A-B\right)+\left(C-D\right) +\left[\left(C-D\right) +\left(\pi-2B\right) + \left(C-D\right) \right] + \dots +\left(C-D\right)+\left(A-B\right)$, where the dots indicate the number of times the square-bracketed term must be included in order to account for the number of internal reflections for the appropriate order $k$ of the rainbow. Overall, then, we have the following so-called deflection functions
\begin{equation}
\vartheta_k^\beta=2\left(A-B\right)+2\left(k+1\right)\left(C-D\right) + k\left(\pi-2B\right),\label{eq:thetabeta}
\end{equation}
which are the analog of the natural rainbow's deflection functions $\vartheta_k^{0}=2\left(A-B^0\right) + k\left(\pi-2B^0\right)$, although for the natural rainbow the internal angle is $B^0=\arcsin\left(\sin\left(A\right)/n_w\right)$ and depends on the refractive index $n_w$ of water only. Similarly to Eq.\ \eqref{eq:thetabeta}, the related $\alpha$-ray's deflection is given by $\vartheta_k^\alpha=2\left(A-B\right)+2\left(C-D\right) + k\left(\pi-2D\right)$. Each of these ray-path types undergoes a minimum angular deviation, and correspondingly, each ray path will have its corresponding rainbow caustic. Although these results do not admit analytic expressions for the angles of minimum deviation, i.e.\ the rainbow angles, using the above expressions these may be derived numerically. To do so, one seeks the root $\underline{A}$ of the first derivative of the deflection function $\partial \vartheta_k^{\dots}/\partial A|_{A=\underline{A}}=0$ and inserts this incidence angle (which specifies the so-called Cartesian ray) back into the expression for the ray's deviation to get the minimum deviation angle $\underline{\vartheta}_{k}^{\dots}=\vartheta_{k}^{\dots}\left(\underline{A}\right)$. The underscore refers to the fact that we are dealing with a function minimum. Alternatively, approximation formulas may be derived: For the primary bow ($k=1$) these have been given in the form of $\underline{\vartheta}_k^{\dots}\approx \underline{\vartheta}_k^{0}  + \delta \underline{\vartheta}_k^{\dots}$,\cite{Lock1994}
\begin{eqnarray}
\delta \underline{\vartheta}_1^\alpha&=& \frac{2\delta}{r_i}\left[\left(\frac{4-n_w^2}{n_w^2+3n_g^2-4}\right)^{1/2}-\left(\frac{4-n_w^2}{n_w^2-1}\right)^{1/2}\right]\label{eq:thetbetaApprox}\\
\delta \underline{\vartheta}_1^\beta&=& \frac{2\delta}{r_i}\left[2\left(\frac{4-n_w^2}{n_w^2+3n_g^2-4}\right)^{1/2}-\left(\frac{4-n_w^2}{n_w^2-1}\right)^{1/2}\right]\nonumber
\end{eqnarray}
with $\delta = r_o-r_i$ being the wall-thickness and $\delta /r_i$ the expansion parameter employed in the series expansions leading to Eq.\ \eqref{eq:thetbetaApprox} (see Appendix \ref{AppendixApprox}). For the second-order rainbow ($k=2$), we state it here for the $\beta$-ray only:
\begin{equation}
\delta \underline{\vartheta}_2^\beta= \frac{2\delta}{r_i}\left[3\left(\frac{9-n_w^2}{n_w^2+8n_g^2-9}\right)^{1/2}-\left(\frac{9-n_w^2}{n_w^2-1}\right)^{1/2}\right]\label{eq:thetbetaApprox2}
\end{equation}
These approximations perform well for low refractive index contrasts and small wall-thicknesses $\delta/r_i \ll 1$. It should be noted, however, that interference effects which theoretically occur at very small coating thicknesses have not been considered in the geometrical optics treatment.\cite{Lock1994} 

From the above, one finds that for the typical situation with $n_g>n_w$ the effect of the additional refractions relative to the natural rainbow situation is a \textit{decreased total deviation angle for all rainbow-orders} $k$. Mathematically, this means that all deviation perturbations are negative ($\delta \underline{\vartheta}<0$) for both the primary as well as the secondary rainbows, as well as for both types of rays. Since $|\delta \underline{\vartheta}_k^\alpha|>|\delta \underline{\vartheta}_k^\beta|$, the separation of the (weaker) $\alpha$-ray from the natural rainbow theory value is larger.

\textit{Twinning of the primary rainbow} refers to the above described appearance of two close-by rainbows near $\underline{\vartheta}_1^0$: A bright one for the $\beta$-rays, and a dim one for the $\alpha$-rays. Their angular separation is small and proportional to the quantity $\delta /r_i$ as seen by the above equations \eqref{eq:thetbetaApprox}. This twinned rainbow phenomenon may be misinterpreted (were it not for their identical color sequence) as a primary and secondary rainbow, although we have seen that it is only two variations of the primary rainbow.

\begin{figure}[t]
\begin{center}\includegraphics [width=1.0\columnwidth]{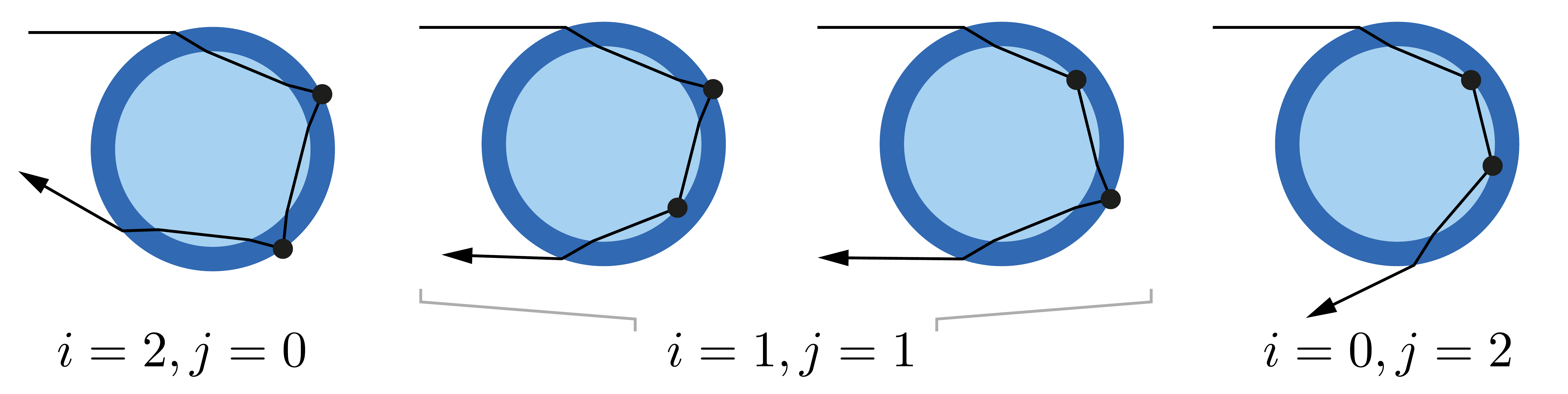}\end{center} %
\caption{Illustration of the concept of degenerate rays relevant for the multiplicity of splitting. For the $k=2$ second-order rainbow, $2^k=4$ different paths are possible, while only $k+1=3$ deflection functions are distinct. \textit{Consequently, a three-fold splitting occurs for the secondary rainbow}.\label{Fig_Degeneracy}}
\end{figure}

The actual secondary rainbow's components shall be considered next in a more general setting. To do so, we first note that the number of possible ray paths increases with the rainbow-order $k$: For each internal reflection, an $\alpha$-like inner reflection or a $\beta$-like outer reflection may occur, such that $2^k$ different paths are available. However, the total deflection only depends on the overall number $j$ of inner reflections and the overall number $i$ of outer reflections (excursions through the wall medium).\cite{Lock2012} That is, the particular succession or ordering of these types of reflections is irrelevant. This may be seen by considering the deflection function which reads:
\begin{equation}
\vartheta_k^{i,j}=2\left(A-B\right)+2\left(C-D\right) + \left\{ \begin{array}{l}
i\left[2\left(C-D\right) + \left(\pi-2B\right)\right]\\
j\left(\pi-2D\right)\\
\end{array}\right.\nonumber
\end{equation}
with $i+j=k$ being the total number of internal reflections. One therefore finds only a $\left(k+1\right)$-fold splitting corresponding to ray paths with distinct deflection functions among the $2^k$ different ray paths. Several ray paths are thus degenerate in the sense that their deflection functions coincide,\cite{Lock2012} see Fig.\ \ref{Fig_Degeneracy}(middle). For $k=2$, a three-fold splitting of the secondary rainbow is thus expected in theory. The above equation contains the previously discussed deflection functions $\vartheta_k^{\alpha}=\vartheta_k^{0,k}$ and $\vartheta_k^{\beta}=\vartheta_k^{k,0}$ as special cases. Each outer reflection causes a total deviation larger by an angle $\vartheta_k^{i,j}-\vartheta_k^{i-1,j+1}=2\left(C-B\right)>0$ in comparison with an inner reflection, see Fig.\ \ref{Fig_Sketch}. This implies that each inner reflection moves the corresponding ray further apart (with a correspondingly larger negative perturbation) from the deviation of an analogous ray for the natural rainbow, i.e.\ $\vartheta_k^{i-1,j+1} < \vartheta_k^{i,j} < \vartheta_k^0$, cf.\ also Fig.\ \ref{Fig_Degeneracy} illustrating the case $k=2$.

\section{Alexander's dark band}
\begin{figure}[bth]
\begin{center}\includegraphics [width=1.0\columnwidth]{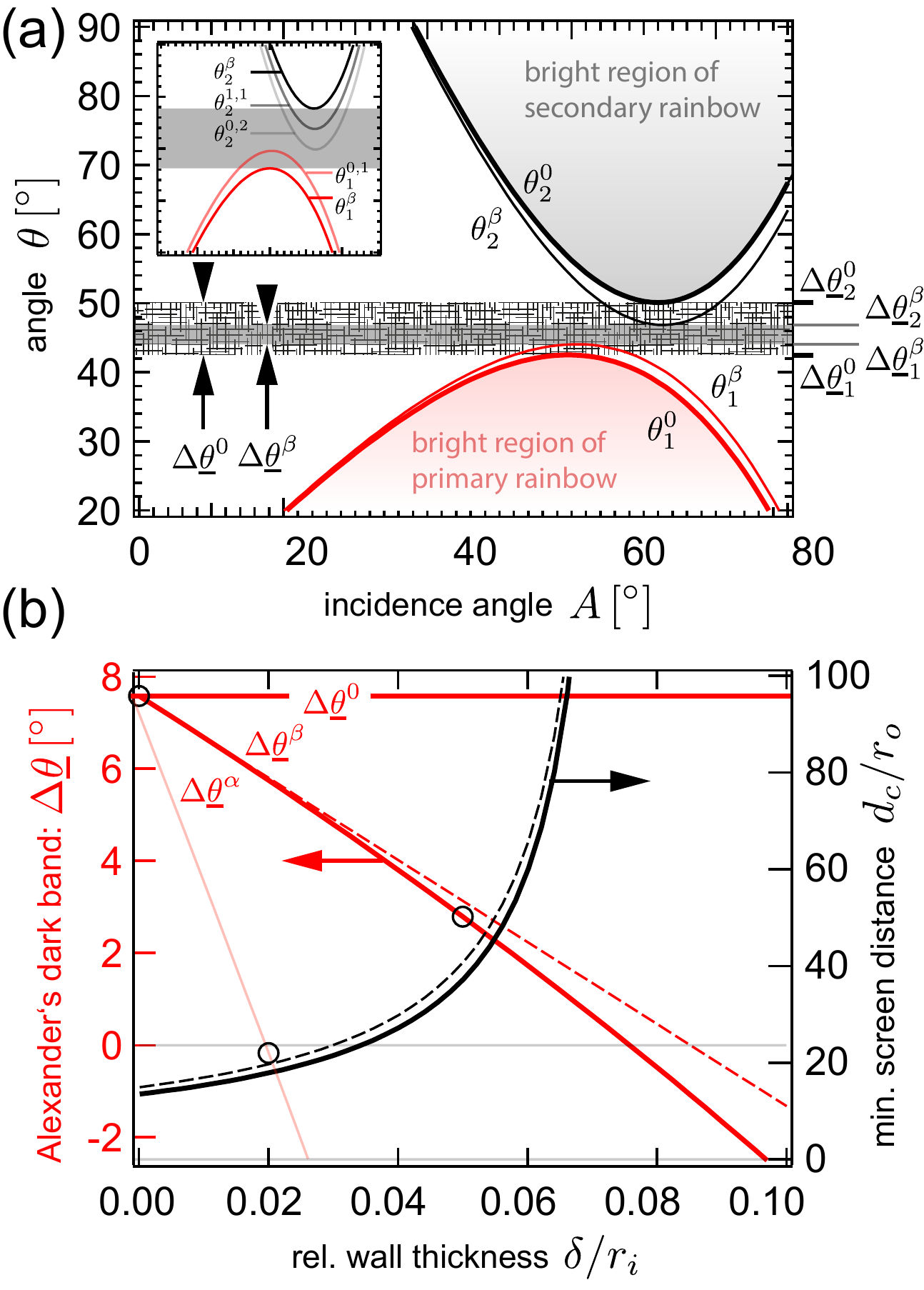}\end{center} %
\caption{\textbf{(a)} \textit{Deflection functions} $\theta^0_{1,2}$ for the first- and second-order natural rainbow (thick lines) as well as $\theta^\beta_{1,2}$ for the flask rainbow (thin lines); $n_w=1.33$ (Appendix \ref{Appendix:nW}), $n_g=1.47$, $\delta/r_i=0.05$. \textit{Inset}: split-structure for $\delta/r_i=0.02$, showing also the faint components $\theta^{0,1}_{1}$, $\theta^{1,1}_{2}$, $\theta^{0,2}_{2}$. \textbf{(b)} \textit{Angular width} (red) of Alexander's dark band for different ratios $\delta / r_i$. The dashed line shows the linear approximation Eq.\ \eqref{eq:ThetaAlexander} using expressions \eqref{eq:thetbetaApprox} and \eqref{eq:thetbetaApprox2}. For the parameters used, it is $\Delta \underline{\theta}^\beta \approx 7.6^{\circ} - 89^{\circ} \times \delta/r_i$. The circles denote the two scenarios of (a). The black curves show the \textit{minimum screen distance} $d_c$ required to see Alexander's dark band. The dashed line is the approximation Eq.\ \eqref{eq:CritDist}, the solid line is the actual distance derived in Appendix \ref{Appendix:exactdc}.\label{Fig_Alexander}}
\end{figure}
Considering only deviations within a range of $0-180^{\circ}$, i.e.\ discarding full revolutions (multiples of $2\pi$) of a given ray and thereby reducing angular coordinates to absolute values instead of signed deviations, we take 
\begin{equation}
\theta = |\pi- \left(\vartheta \,{\rm mod}\, 2\pi\right)|.
\end{equation}
This means, for the primary rainbow we will be dealing with $\underline{\theta}_1=\pi-\underline{\vartheta}_1$ (cf.\ Fig.\ \ref{Fig_Sketch}, inset), whereas for the secondary we will have $\underline{\theta}_2=\underline{\vartheta}_2-\pi$. Accordingly, the effect of the decreased total deviation angle upon increasing wall-thickness is a shift towards larger angular coordinates for the minimum deviation angle of the primary, and a shift towards smaller angles for the secondary rainbow, see Fig.\ \ref{Fig_Alexander}(a). This bears consequences on the dark region between both rainbows, i.e.\ Alexander's dark band. We will first consider the bright $\beta$-components of both the primary and the secondary rainbows: \textit{The angular width} 
\begin{equation}
\Delta \underline{\theta}^\beta =\underline{\theta}_2^{\beta}-\underline{\theta}_1^{\beta}\approx \Delta \underline{\theta}^0 + (\delta \underline{\vartheta}_2^\beta + \delta\underline{\vartheta}_1^\beta)\label{eq:ThetaAlexander}
\end{equation}
\textit{of Alexander's dark band decreases (approximately linearly) for increasing wall-thicknesses} $\delta/r_i$. This is shown by the red solid (approximation: red dashed) line in Fig.\ \ref{Fig_Alexander}(b), where the dark band finally disappears for the given parameters at $\delta / r_i = 0.075$. We will see a corresponding transition in ray tracing in section \ref{sec:RayTracing}.

What we ignored so far was the role of the less bright splitting components of both the primary and the secondary rainbow. As the $\alpha$-like components are more sensitive to the wall effect, they lie \textit{within} Alexander's dark band discussed above, see the inset of Fig.\ \ref{Fig_Cartesian}(a). Defining the innermost dark region between all splitting components of each the rainbow via $\Delta \underline{\theta}^\alpha = \underline{\theta}_2^{0,2}-\underline{\theta}_1^{0,1}$, one finds that, strictly speaking, Alexander's dark band already disappears for wall-thicknesses larger than $\delta /r_i>0.020$, see the light red line in Fig.\ \ref{Fig_Alexander}(b). However, the intensities $I_{i,j}$ of the weaker splitting components of each rainbow order are in fact very low. Considering the ratio of the relevant polarization-averaged Fresnel coefficients,\cite{Whitaker1974,Adam2002} one finds approximately the following intensity ratios: $I_{1,0}/I_{2,0} \sim 3:1$ (primary rainbow : secondary rainbow), $I_{1,0}/I_{0,1} \sim 20:1$ ($\beta$ : $\alpha$ for primary rainbow) and $I_{2,0}:I_{1,1}:I_{0,2} \sim 1500:50:1$ for $\delta/r_i \approx 0.030$ (and similar values for other wall-thicknesses between $0<\delta/r_i < 0.1$). Therefore, the above restriction is not really of any practical concern and it suffices to consider the dark band $\Delta \underline{\theta}^\beta$, Eq.\ \eqref{eq:ThetaAlexander}, between the most intense components. Incidentally, as we have seen before, these are also those which, in terms of angular coordinates, lie closest to the natural phenomenon's rainbow orders.

\section{Critical distance of the screen\label{sec:RayTracing}}
We continue by looking at the same phenomena as before, but this time from the perspective of ray tracing. In particular, we will be concerned with the distance beyond which Alexander's dark band emerges.

\begin{figure}[bth]
\begin{center}\includegraphics [width=1.0\columnwidth]{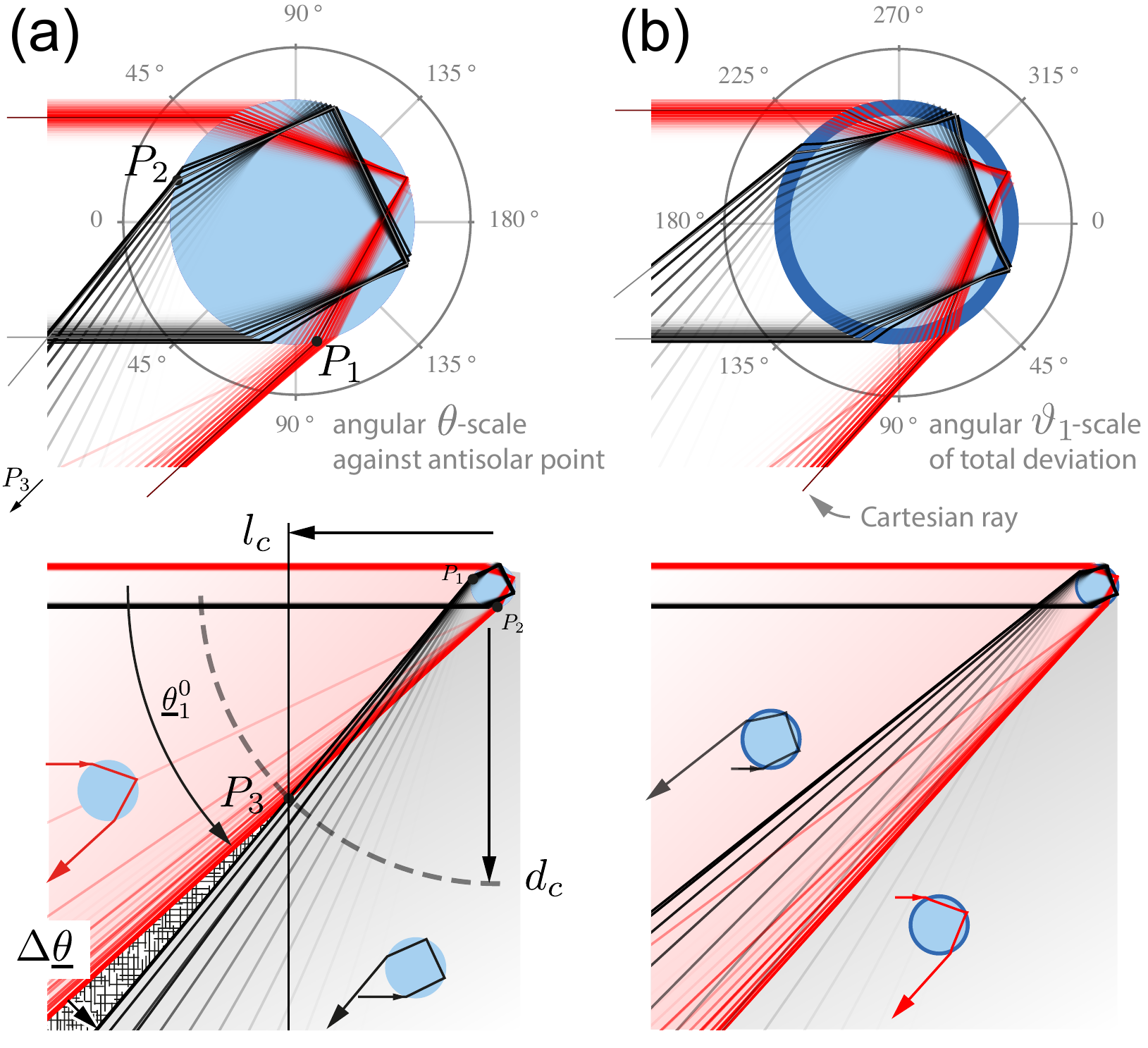}\end{center} %
\caption{\textbf{(a)} Ray tracing for the primary and secondary rainbow rays for a \textit{water sphere} with $n_w=1.33$. (For better clarity, rays have been transparency-coded according to their proximity to the Cartesian ray's impact parameter. The actual intensity in angular space is due to the relative abundance of rays with deviations around the minimum deviation angle $\underline{\theta}_k$.) When $\Delta \underline{\theta}^\beta>0$, the points $P_1$ ,$P_2$ and $P_3$ approximately define an isosceles triangle. Eq.\ \eqref{eq:CritDist} may then be used to estimate the critical distance $d_c$ beyond which the secondary rainbow emerges from within the bright region of the primary rainbow and Alexander's dark band begins to be observable. \textbf{(b)} Same as in (a), but for a \textit{water-filled spherical flask} with $n_g=1.47$ and $\delta/r_i = 0.15$ (i.e.\ $r_i/r_o\approx 0.86$). Here, the secondary rainbow lies within the primary rainbow and $\Delta \underline{\theta}^\beta <0$, i.e.\ Alexander's dark band does not exist.\label{Fig_Cartesian}}
\end{figure}

\begin{figure}[hbt]
\begin{center}\includegraphics [width=1.0\columnwidth]{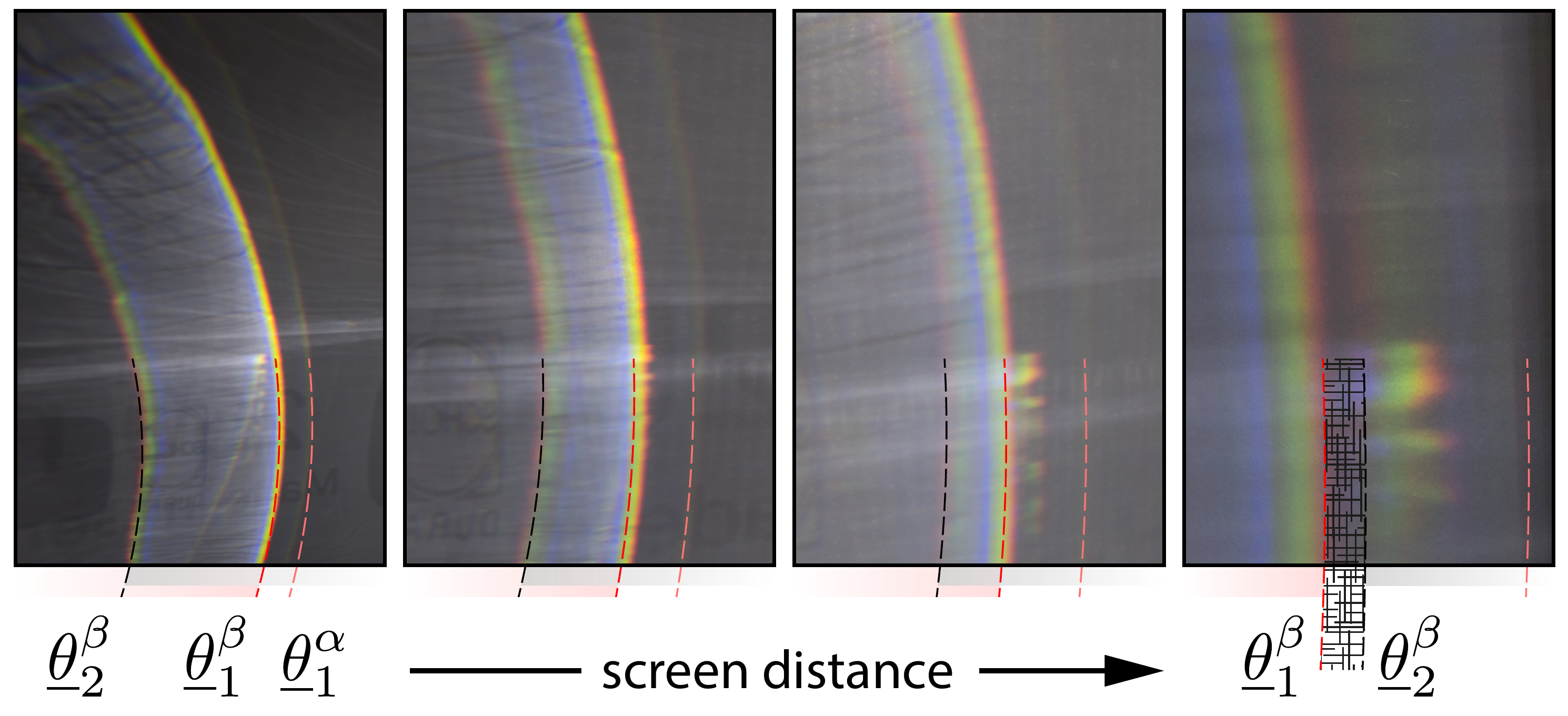}\end{center} %
\caption{\textbf{(a)} Rainbow projections at around $\underline{\theta}_1^\beta$ for increasing screen distances (left to right) for the \textit{water-filled flask} of Fig.\ \ref{Fig_Exp}. The screen size was $20\times 30 \,\rm cm$. The last image corresponds to a distance $1.50\,{\rm m} > d_c= 1.22\,\rm m$ (cf.\ Appendix \ref{Appendix:exactdc}) where Alexander's dark band has finally emerged.\label{Fig_Experiment}}
\end{figure}

Figure \ref{Fig_Cartesian}(a),(b) shows the distinguishing characteristic between the situation in which a dark band exists, and the situation where it has disappeared: In the former case (a), e.g.\ for the natural phenomenon, rays for the first and the second-order rainbows which enter the sphere from two opposing sides will be deflected into the same half-space. The bundleling of rays close to the caustic signifies the high intensity perceived under the rainbow angles. One observes that beyond a certain distance the rays open up a space between them where neither first-order nor second-order rainbow rays are deflected into (the criss-cross-shaded area). This is Alexander's dark band in the near-field, visible here as a wedge of angle $\Delta \underline{\theta}$ (for a similar plot see Ref.\ \cite{Walker1980}). In contrast, in the latter case (b), i.e.\ for thick walls, the Cartesian rays for $k=1,2$ emerging from opposing sides of the sphere no longer converge to cross and eventually open up a dark region between them, and Alexander's dark band disappears.
The relevant Cartesian rays for the first and the second-order rainbows emerge from nearly opposite sides of the sphere, see points $P_1$ and $P_2$ in Fig.\ \ref{Fig_Cartesian}. We will exploit this fact to derive an approximation for the distance $d_c\gg r_o$ where they meet (if they do) and beyond which they open up Alexander's dark band (i.e.\ if it exists). Considering that $\underline{\theta}_2 > \underline{\theta}_1$, they span (approximately) an isosceles triangle with two sides of length $d_c$ and one of length $\sim 2 r_o$, with an apex at point $P_3$ containing an angle corresponding to Alexander's dark band width $\Delta \underline{\theta}$. Therefore, we may find a good approximation for the distance $d_c$ at which both Cartesian rays intersect from 
\begin{equation}
\sin\left(\frac{\Delta \underline{\theta}}{2}\right)\approx \frac{r_o}{d_c}\label{eq:CritDist}.
\end{equation}
An exact derivation of $d_c=|P_3|$ is given in Appendix \ref{Appendix:exactdc}. The significance of this distance being that Alexander's dark band appears only if observed / projected at a distance larger than $d_c \approx r_o \sin\left(\Delta \underline{\theta}/2\right)$. In front of that distance, the secondary rainbow is still inside the bright region of the primary rainbow, see Figs.\ \ref{Fig_Exp}, \ref{Fig_Cartesian}, \ref{Fig_Experiment} and also Fig.\ \ref{Fig_CartesianAcrylic}(c), such that instead a \textit{bright band} appears.

Using the typical geometry of Florence's rainbow demonstration experiment, this translates into a critical orthogonal minimum distance of the plane screen of $l_c \approx d_c\cos\left(\underline{\theta}_1\right)$, see Fig.\ \ref{Fig_Cartesian}. For a water drop (without coating), the critical distance becomes $d_c\approx 15 r_o$ ($l_c\approx 11 r_o$) and is easily achieved for a liquid drop experiment where $r_o\sim 1-2\,\rm mm$.\cite{Walker1975,Walker1980,Tammer1998} In the Florence flask experiment, this minimum distance becomes macroscopic and less easy to realize. Even worse, the effect of a finite vessel wall-thickness decreases Alexander's dark band width $\Delta \underline{\theta}$, and hence increases approximately in inverse proportion the required distance according to $d_c\propto 2r_o/\Delta \underline{\theta}$.


For a typical 250mL-Florence flask with an outer diameter of $\varnothing = 2r_o=85\,\rm mm$ and a wall-thickness of $\delta=1.7\,\rm mm$ ($\delta/r_i \approx 0.042$) the critical minimum distance is thereby increased from $d_c\approx 64\,\rm cm$ ($l_c \approx 47\,\rm cm$, both assuming no wall) to $1.34\,\rm m$ ($0.97\,\rm m$), owing to the reduction of Alexander's dark band width from $\Delta \underline{\theta}^0=7.6^{\circ}$ to $\Delta \underline{\theta}^\beta = 3.6^{\circ}$ through the effect of the finite wall-thickness (slightly smaller exact values are reported at the end of Appendix \ref{Appendix:exactdc}). Already at $\delta=2.5\,\rm mm$ wall-thickness this distance increases to $3\,\rm m$, and thereafter rapidly increases until it diverges at around $\delta=3.0\,\rm mm$. That is, for a 250mL Florence flask, Alexander's dark band can be observed only for wall-thicknesses less than this value. Also, a full circular double-rainbow will potentially require an impractically large screen ($\sim 2\times 2\,\rm m$) in the usual setup (cf.\ Fig.\ \ref{Fig_Exp}).\cite{Greenslade2982,Johnson,Minnaert1993,Ticker,Bohren1980,Calvert,Harvard} However, parts of Alexander's dark band may be observed if a screen is placed at a sufficient distance, see the right-most photograph in Fig.\ \ref{Fig_Experiment}.

\section{More ray-types?\label{sec:MoreRayTypes}}
This short section aims to briefly show that the ray types considered thus far are sufficient when discussing the experiment. Only for unpractically large wall-thicknesses entirely new ray trajectories begin to appear:\cite{Lock1994} For instance, when $r_i/r_o<1/n_w \approx 0.75$,\cite{Wang2014} total internal reflection at the inner glass-water interface can then prevent the rainbow rays to form at all. In this case, the wall acts as a waveguide.\cite{Lock1994} For thicker walls with $r_i/r_o < 1/n_g \approx 0.68$, rays which only transit the glass-material may occur. When the glass wall becomes thicker still, eventually a glass-rainbow emerges when these rays experience a minimum deviation. Details are given in Appendix \ref{AppendixThick}. 

Since normal lab-supply flasks are rather thin-walled, say $r_i/r_o > 0.9$, the above phenomena (while being interesting on their own) had not to be considered with regard to their impact on the experiment.

\section{Variations}
\begin{figure}[b]
\begin{center}\includegraphics [width=1.0\columnwidth]{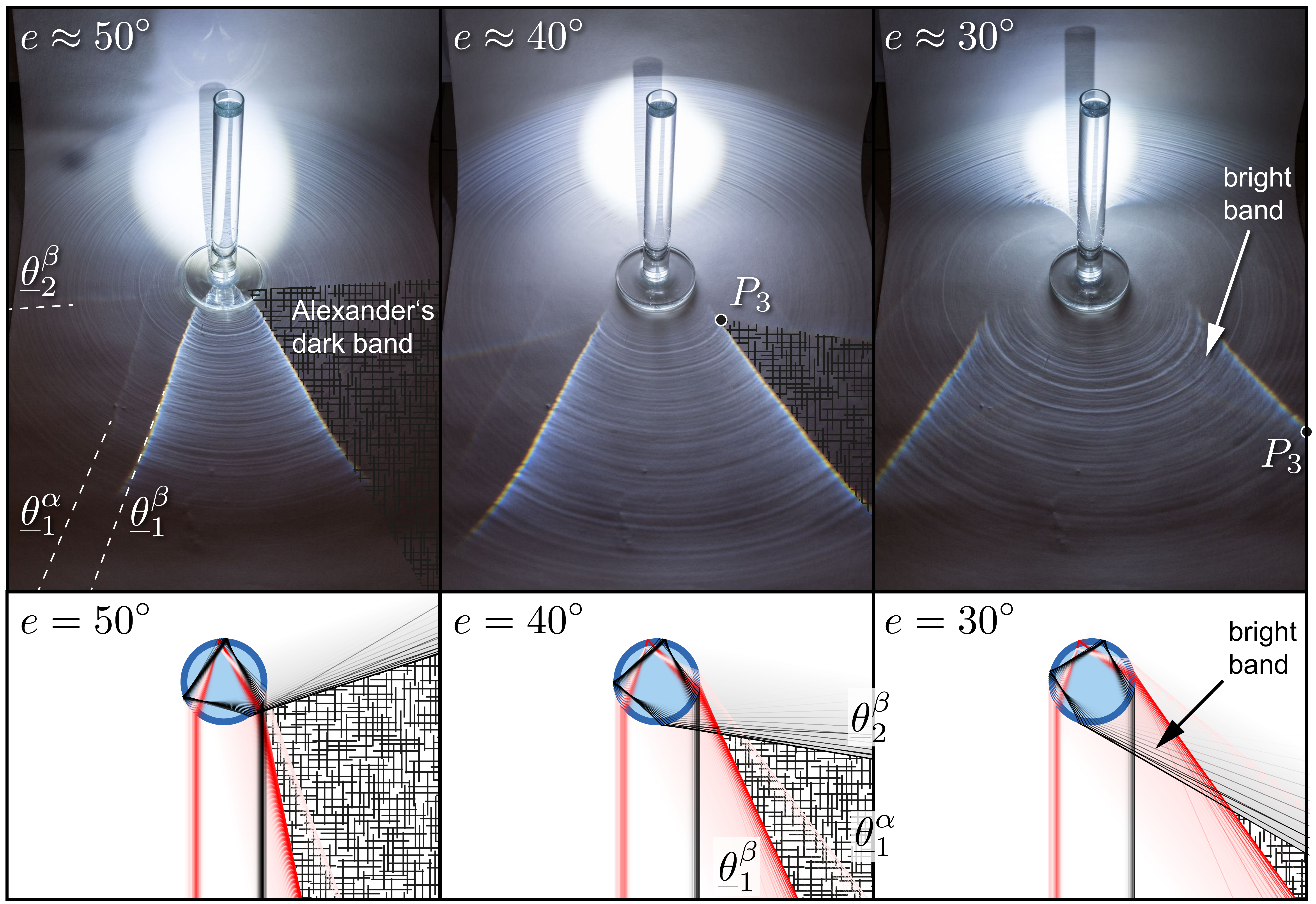}\end{center} %
\caption{A cylindrical water-filled vase ($\varnothing=3.2\,\rm cm$, height: $30\,\rm cm$, $\delta/r_i=0.19$) placed on the floor and illuminated using a focused flash light. The screen (photographed at an angle!) forms an "L" as it rests against the wall, thus showing the projected shadow of the vase. The light's inclination angle $e$ is lowered from left to right, thereby tuning Alexander's dark band. Corresponding ray tracings are shown below each scenario. Light red rays show the $\alpha$-rays for $k=1$. For $e=\left\{50^{\circ},40^{\circ},30^{\circ}\right\}$ one finds $\{\underline{\theta}_1^\alpha,\underline{\theta}_1^\beta,\underline{\theta}_2^\beta,\Delta \underline{\theta}^\beta\}=\left\{21^{\circ},12^{\circ},108^{\circ},96^{\circ}\right\}, \left\{37^{\circ},25^{\circ},80^{\circ},55^{\circ}\right\}, \left\{51^{\circ},36^{\circ},60^{\circ},24^{\circ}\right\}$.\label{Fig_Vase}} 
\end{figure}

This article's considerations also apply to rainbow demonstrations using a cylindrical glass of water resting on a level support, e.g.\ on a table.\cite{Selmke2016} Only in this case the incidence angle $e$ of the light towards the table (i.e.\ the base of the cylinder) requires the use of Bravais' index of refraction for inclined rays. One can show that the transformations $n_g,n_w\rightarrow n_g',n_w'$ according to Eq.\ (7) of Reference [\!\!\citenum{Selmke2016}] should be applied. Similar to the ray paths contributing to the parhelic circle halo, the inclination is not altered in the overall transit via mutually compensating refractions.\cite{Selmke2015} Since the inclination increases the (primed) effective refractive indices, the rainbow angles and the dark band projected on the table may be tuned and in particular broadened by this mechanism. The wall-effect becomes significant if kitchen experiments are made using cylindrical vases which typically have rather thick walls, see Fig.\ \ref{Fig_Vase}.

\begin{figure}[bth]
\begin{center}\includegraphics [width=1.0\columnwidth]{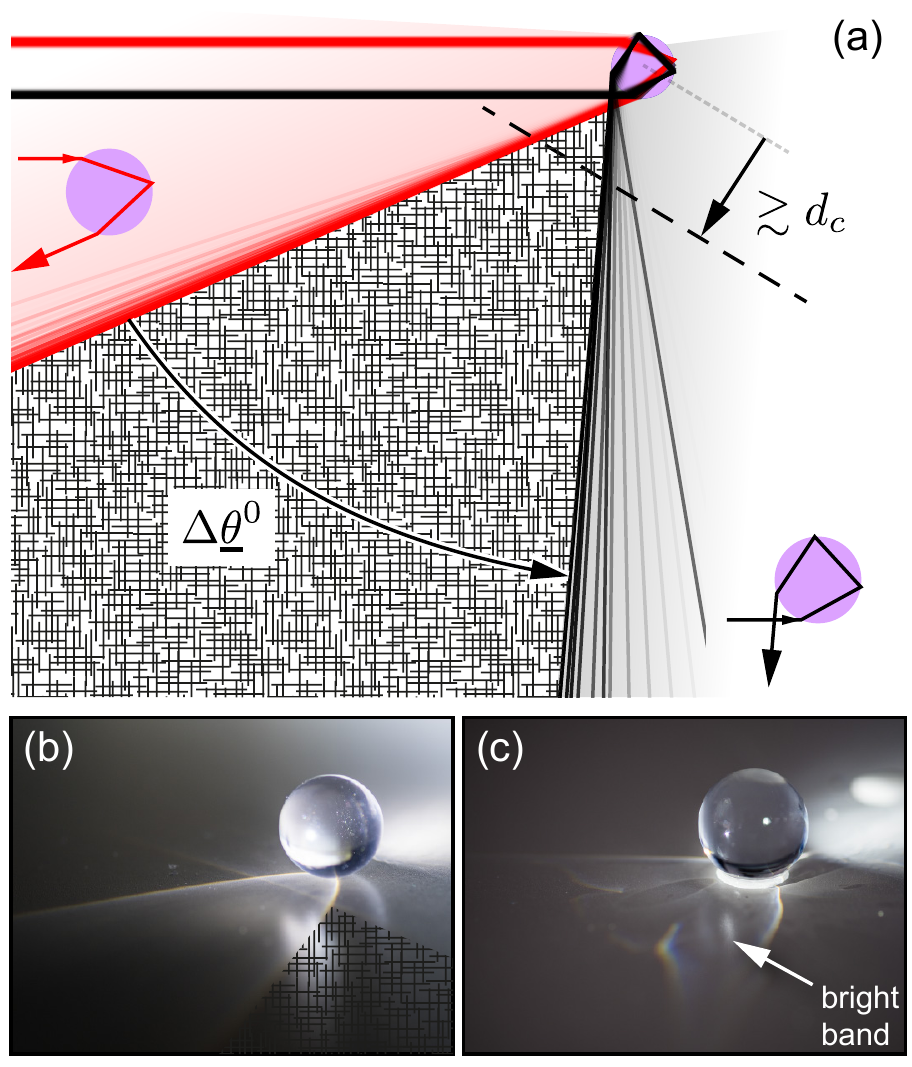}\end{center} %
\caption{Uncoated spheres: \textbf{(a)} ray tracing as in Fig.\ \ref{Fig_Cartesian} but for an \textit{acrylic sphere} with $n=1.49$. Alexander's dark band will be best visible when a screen (dashed line) is placed beyond but close to $d_c$, e.g.\ $\sim$ a sphere diameter away from the surface of the sphere. The photos show spheres ($\varnothing \approx 10\,\rm mm$) resting on a white surface (effectively: $d\in \left[ r_o,+\infty \right]$) and illuminated by parallel (to the surface) white light from a focusable flashlight: \textbf{(b)} Acrylic sphere. \textbf{(c)} Magic (plant) growing jelly ball / water bead with $\sim n_w$ forming a bright band in the near-field, cf.\ Fig.\ \ref{Fig_Cartesian}(a).\label{Fig_CartesianAcrylic}} 
\end{figure}

As mentioned in the introduction, another possible remedy for the particularities associated with the flask experiment is to use a solid glass or acrylic sphere\cite{Harvard,Selmke2015} (or disc\cite{Casini2012}) and to place the screen at a very close distance. While such an experiment will not give a faithful representation of the large far-field band width ($\Delta\underline{\theta}^0=61^\circ$) or the (non-natural) rainbow angles ($\underline{\theta}_1^0=24^\circ$, $\underline{\theta}_2^0=85^\circ$) for a refractive index of $1.49$, it will nicely show a double rainbow (better: glassbows) and Alexander's dark band beyond but close to the critical distance $d_c=1.66 r_o$ or $l_c=1.04r_o$, see Fig.\ \ref{Fig_CartesianAcrylic}(a)(b).


\section{Conclusion}
The intuitively accessible Florence flask rainbow demonstration experiment has a firm place in optics classes around the world, either at the high-school level or at a graduate level at universities. The unavoidable finite wall-thickness of the artificial raindrop (spherical vessel) causes several changes with respect to the natural phenomenon: The rainbow angles are slightly modified from the natural rainbow's theory. This, in turn, narrows down Alexander's dark band. Additional possible ray paths in this geometry lead to weaker split-components of all rainbow orders which further impede an easy identification of rainbow orders and a truly dark band. In any case, we found that a critical minimum screen distance is required for a projection of Alexander's dark band. Detailed considerations along the lines put forward within this article may be used to either guide the demonstration experiment's design or to quantitatively analyse this popular rainbow experiment (or similar ones).

\newpage
\appendix
\section{The angle $C$\label{AppendixAngleC}}
To find the angle $C$,\cite{Lock1994,Wang2014} we use the cosine theorem twice in the triangle of Fig.\ \ref{Fig_Sketch} with sides $r_o$, $r_i$ and the first segment between the outer and the inner surface of length $p$ (with internal angles $B$, $C-B$ and $\pi-C$):
\begin{eqnarray}
p^2&=&r_o^2+r_i^2-2r_o r_i \cos\left(C-B\right),\label{eq:Cos1}\\
r_i^2&= &r_o^2+p^2 - 2r_o p \cos\left(B\right).\label{eq:Cos2}
\end{eqnarray}
These are two equations with two unknowns, the length $p$ and the angle $C$. We may thus proceed to solve for $C$. First, Eq.\ \eqref{eq:Cos1} is solved for $C$, %
\begin{equation}
C=B + \arccos\left(\left[r_o^2+r_i^2 - p^2\right]/2r_o r_i\right).\label{eq:C}
\end{equation}
The unknown length $p$ in this expression may be found from the appropriate solution to the quadratic equation \eqref{eq:Cos2}, $p=r_o \cos\left(B\right) - \sqrt{r_o^2 \cos^2\left(B\right)-\left(r_o^2-r_i^2\right)}$. Unfortunately, no significant further simplification seems possible, which is ultimately the reason why no closed-form expressions for the rainbow angles can be found.

%

\section{Approximation formulas for $\delta\underline{\theta}_k^\beta$\label{AppendixApprox}}
The approximation expressions $\delta \underline{\vartheta}_k^\beta =\underline{\vartheta}_k^\beta - \underline{\vartheta}_k^0$ may be obtained in the following way: First, the angle $C$, Eq.\ \eqref{eq:C}, and its corresponding angle $D$ are expanded to first-order in $\delta / r_i$ using a computer algebra system, e.g.\ $C=B + \tan\left(B\right) \delta/r_i + \mathcal{O}\left(\delta^2/r_i^2\right)$. The correction to the rainbow theory value is then $\delta \vartheta_k^\beta=2\left(k+1\right)\left(C-D\right)+2\left(k+1\right)\left(-B+B^0\right)$, in which the former term contains the additional deviations related to transits through the wall, and the latter term corrects the deviations $2\left(A-B\right)$ and $k\left(\pi-2B\right)$ in the sum of deviations in Eq.\ \eqref{eq:thetabeta} for using $n_g$ instead of $n_w$. Thereafter, the unperturbed analytical rainbow angle expressions\cite{Walker1975,Casini2012,Adam2002} for $A=\underline{A}$ (corresponding to $\underline{\vartheta}_k^0$) are inserted and the approximations for $C$ and $D$ used. This yields $\delta \underline{\vartheta}_k^\beta=2\delta/r_i \times \left[\dots\right]$, with $\kappa=k+1$ and $\left[\dots\right]=$
\begin{equation}
\kappa\left(\frac{\kappa^2-n_w^2}{n_w^2+\left(\kappa^2-1\right)n_g^2-\kappa^2}\right)^{1/2}-\left(\frac{\kappa^2-n_w^2}{n_w^2-1}\right)^{1/2}.\nonumber
\end{equation}
The given approximation formulas Eq.\ \eqref{eq:thetbetaApprox}\cite{Lock1994} and \eqref{eq:thetbetaApprox2} follow with $k=1$ and $k=2$. 


\section{Refractive index of water\label{Appendix:nW}}
\begin{figure}[htb]
\begin{center}\includegraphics [width=1.0\columnwidth]{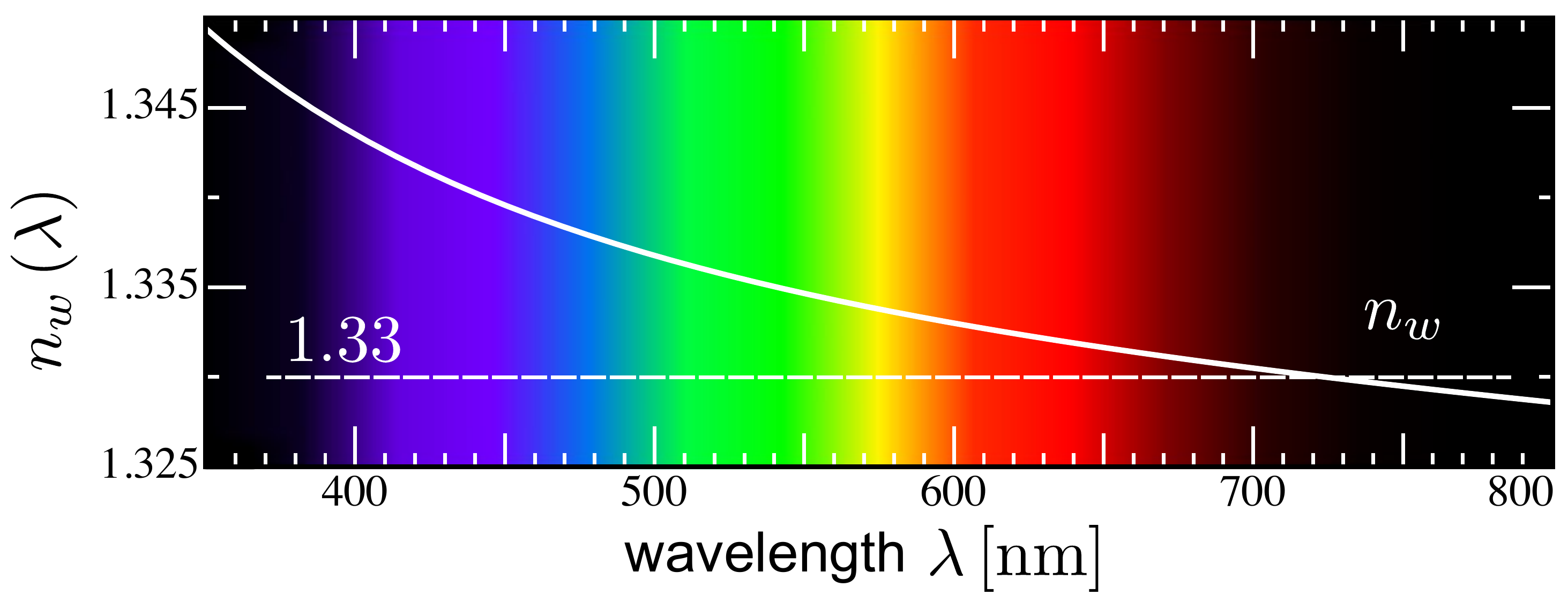}\end{center} %
\caption{The refractive index of (destilled) water $n_w$ for a temperature of $20^{\circ}\rm C$ and standard atmospheric pressure as a function of the wavelength $\lambda$ of light.\cite{Masumura2007} The value $1.33$ (dashed line) was used for all calculations in this article.\label{Fig_nW}}\end{figure}

Throughout the article we have used the value $n_w=1.33$.\cite{Nussenzveig1977} This corresponds to water's refractive index at the red end of the visible spectrum, see Fig.\ \ref{Fig_nW}. Since the given dispersion signifies that red light is refracted less than blue light, we have therefore described the angular positions of also the chromatic minimum of deviations, i.e.\ the truly minimum deviation angles across colors, and thereby the outer limits of the rainbows discussed.

\section{Exact values of $d_c$ and $l_c$\label{Appendix:exactdc}}
\begin{figure}[h]
\begin{center}\includegraphics [width=1.0\columnwidth]{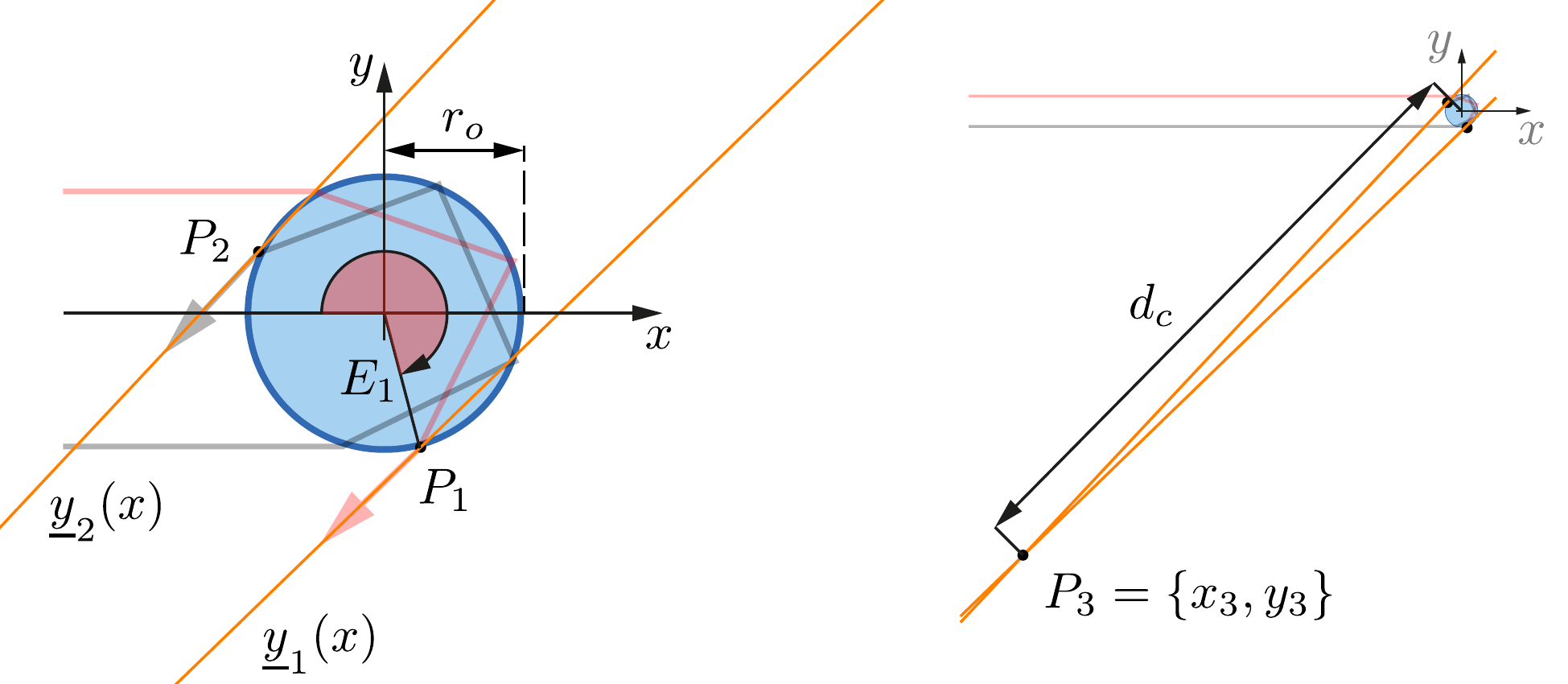}\end{center} %
\caption{\textbf{(a)} Geometry for $\delta/r_i=0.05$ showing exit points $P_1$ and $P_2$. \textbf{(b)} Intersection point $P_3$ of $\underline{y}_1\!\left(x\right)$ and $\underline{y}_2\!\left(x\right)$.\label{Fig_P3}}
\end{figure}

The actual distance $d_c$ of the point $P_3=\left\{x_3,y_3\right\}$ to the center of the sphere may be found by writing down the linear equations for the relevant Cartesian exit rays, assuming incidence from the left (negative $x$-direction). Their exit points $P_{1}$ and $P_{2}$ (considering only the bright $\beta$-type rays) have the Cartesian coordinates $\left\{-r_o\cos\left(E_k\right),\pm r_o\sin\left(E_k\right)\right\}$ where the plus sign is for $k=1$ and the negative sign for $k=2$. The angular coordinates may be inferred from Fig.\ \ref{Fig_Sketch} as $E_k=A+2\left(k+1\right)\left(C-B\right)+\left(k+1\right)\left(\pi-2D\right)$, where $A=\underline{A}$ is to be taken as the angle of the Cartesian ray, i.e.\ the incidence angle for which $\partial \theta_k^\beta /\partial A = 0$. The Cartesian ray's linear equation then reads $\underline{y}_k\!\left(x\right)=\tan\left(\underline{\theta}_k\right) \cdot \left(x + \cos\left(E_k\right)\right) \pm \sin\left(E_k\right)$, where again the plus sign is for $k=1$. The solid (orange) lines in Fig.\ \ref{Fig_P3} show both. Equating both linear equations and solving for $x=x_3$, one may find the coordinates of their intersection point $P_3$. Then, $d_c=|P_3|$ and $l_c=|x_3|$. The former has been plotted as the black solid line in Fig.\ \ref{Fig_Alexander}(b). 


%

In the example case discussed in the main text ($2r_o=85 \,\rm mm$, $\delta=1.7\,\rm mm$, $n_w=1.33$, $n_g=1.47$) this corrects our approximation of $d_c\approx 1.34\,\rm m$ down to $1.22\,\rm m$ and $l_c \approx 0.97\,\rm m$ down to $0.85\,\rm m$. The exact no-wall values, i.e.\ for $\delta=0$, are $d_c=13.7 r_o$ and $l_c=9.5 r_o$. Our initial approximation in section \ref{sec:RayTracing} largely relied on the assumption that $|P_1-P_2| \approx 1.8 r_o$ would be close to $2r_o$, which we thus indeed find to be approximately true.

\section{Phenomena for thick walls\label{AppendixThick}}
In this section, the statements of section \ref{sec:MoreRayTypes} in the main text shall be substantiated and visualized.

\begin{figure}[tbh]
\begin{center}\includegraphics [width=1.0\columnwidth]{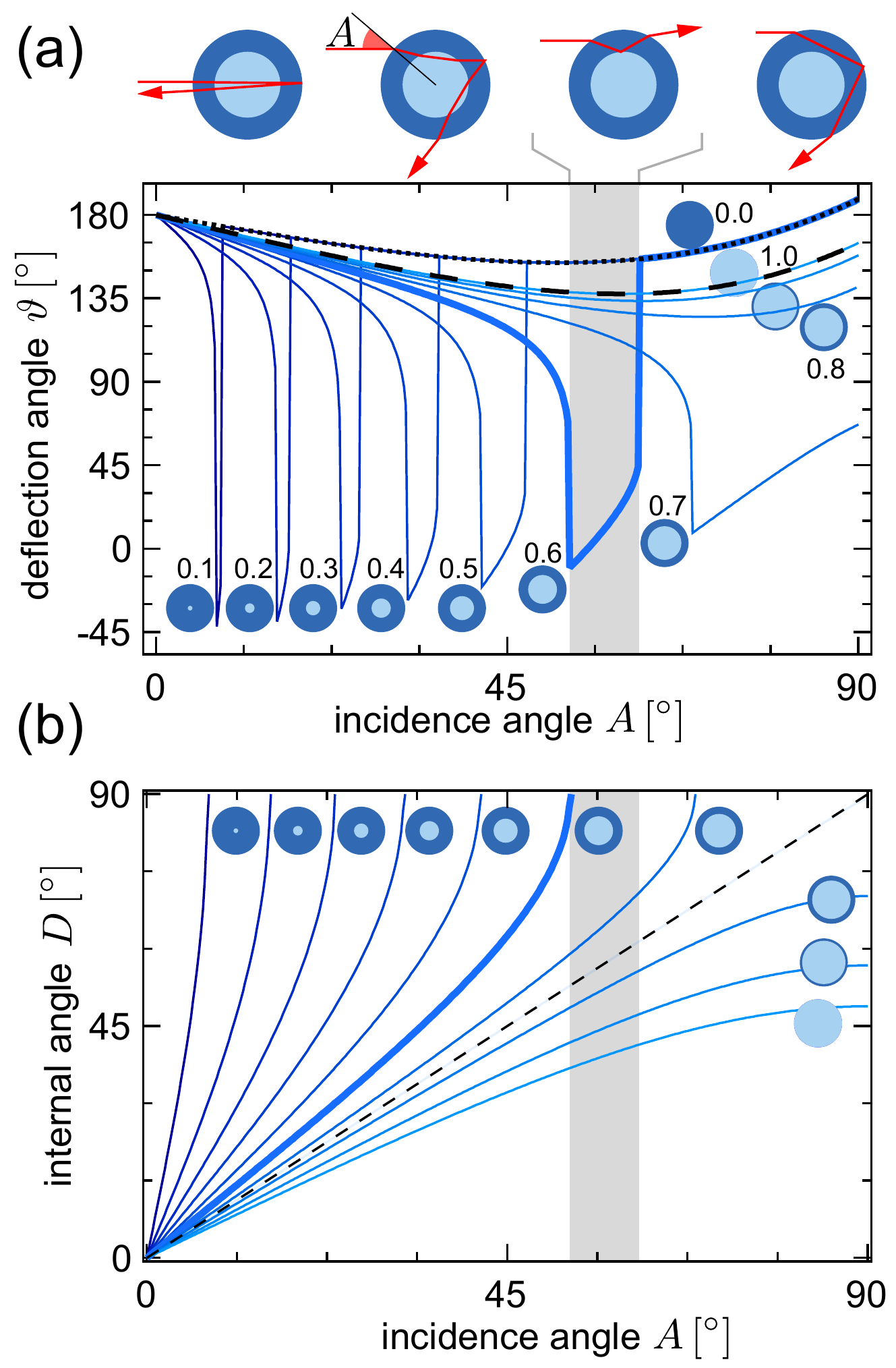}\end{center} %
\caption{\textbf{(a)} Different ratios $r_i/r_o=\left\{0.1,\dots,0.9\right\}=1/\left(\delta/r_i+1\right)$. The ratio $r_i/r_o=0$ represents a glass sphere and its deflection function is the dotted black line. The ratio $r_i/r_o=1$ corresponds to a water sphere, i.e.\ the natural rainbow (black dashed line, cf.\ thick red line in Fig.\ \ref{Fig_Alexander}(a)). The thick solid line emphasizes a single scenario with $r_i/r_o=0.6$. The grey shaded area denotes the parameter space where total internal reflection occurs for this ratio. For curves with $r_i/r_o<1/n_g\approx 0.68$ there is a critical incidence angle above which the curves collapse to the glass rainbow function. For curves with $r_i/r_o<1/n_w \approx 0.75$, total internal reflection occurs and the deflection function changes abruptly. \textbf{(b)} Internal angle $D$ as a function of the incidence angle $A$. TIR sets in when $D\rightarrow \pi/2$. This happens first for $r_i/r_o=1/n_w$ where $A=D$ holds (dashed line).\label{Fig_GraphsThick}}
\end{figure}

Ray paths which stay in the glass medium (i.e.\ do not encounter the water-core) become possible when there exists a ray for which its minimum distance $d_m=r_o \sin\left(B\right)$ within an imaginary solid sphere of glass becomes larger than the inner radius of the actual water sphere inside, $d_m>r_i$. Since $\sin\left(A\right)=n_g\sin\left(B\right)$, the incidence angle for which this occurs would be $A=\arcsin\left(n_g r_i/r_o\right)$, which is only real-valued, i.e.\ accessible if $r_i/r_o < 1/n_g$ or, equivalently, when $\delta/r_i > n_g - 1$. If the inner radius is decreased beyond this point even further, a minimum deviation angle may be reached and a corresponding glass bow which is smaller than the water bow, $\theta_{1,g}^0 < \theta_{1}^0$, emerges. It may happen that neither the $\beta$-rays nor the glass-only rays exhibit a minimum deviation angle where $\partial \theta/ \partial A=0$. The thick line in Fig.\ \ref{Fig_GraphsThick}(a) shows such a case for $r_i/r_o=0.6<1/1.47\approx 0.68$.

Total internal reflection (TIR) sets in when $D=\pi/2$. Since the internal angle $D$ is an increasing function of the incidence angle $A$, the incidence angle is maximal (i.e.\ $A=\pi/2$) when this happens. Thus, setting $\sin\left(B\right)\rightarrow 1/n_g$ accordingly and plugging this into the TIR condition in the form of $\sin\left(C\right)=n_w/n_g$ (see Appendix \ref{AppendixAngleC} for the angle $C$) one can solve for $r_i/r_o$ with a computer algebra system. We find the critical ratio where TIR starts to happen is $r_i/r_o=1/n_w$, or equivalently when $\delta/r_i=n_w - 1$.\cite{Wang2014} Although surprisingly difficult to show, one finds that for $r_i/r_o=1/n_w$ the following equality between the incidence and internal angle holds: $A=D$, cf.\ dashed line in Fig.\ \ref{Fig_GraphsThick}(b). The deflection function for rays which experience total internal reflection and subsequently leave the vessel is $\vartheta^{\rm TIR}=2\left(A-B\right)-\left(\pi-2C\right)$, see for instance the segment of the deflection function in the shaded area in Fig.\ \ref{Fig_GraphsThick}(a).




\end{document}